\documentclass[conference]{IEEEtran}
\IEEEoverridecommandlockouts
\usepackage{cite}
\usepackage{amsmath,amssymb,amsfonts}
\usepackage{algorithmic}
\usepackage{graphicx}
\usepackage{textcomp}
\usepackage{xcolor}
\usepackage{colortbl}
\usepackage{threeparttable}
\usepackage{tcolorbox}
\usepackage{enumitem}
\usepackage{multirow}
\usepackage{comment}

\def\BibTeX{{\rm B\kern-.05em{\sc i\kern-.025em b}\kern-.08em
    T\kern-.1667em\lower.7ex\hbox{E}\kern-.125emX}}
\begin{document}

\title{
{An Empirical Study on Leveraging Images in Automated Bug Report Reproduction}
}

\makeatletter
\newcommand{\linebreakand}{%
  \end{@IEEEauthorhalign}
  \hfill\mbox{}\par
  \mbox{}\hfill\begin{@IEEEauthorhalign}
}
\makeatother

\author{
  \IEEEauthorblockN{Dingbang Wang}
  \IEEEauthorblockA{
    \textit{University of Connecticut} \\
    USA \\
    dingbang.wang@uconn.edu}
  \and
  \IEEEauthorblockN{Zhaoxu Zhang}
  \IEEEauthorblockA{
    \textit{University of Southern California} \\
    USA \\
    zhaoxuzh@usc.edu}
  \and
  \IEEEauthorblockN{Sidong Feng}
  \IEEEauthorblockA{
    \textit{Monash University} \\
    Australia \\
    sidong.feng@monash.edu}
  \linebreakand
  \IEEEauthorblockN{William G. J. Halfond}
  \IEEEauthorblockA{
    \textit{University of Southern California} \\
    USA \\
    halfond@usc.edu}
  \and
  \IEEEauthorblockN{Tingting Yu}
  \IEEEauthorblockA{
    \textit{University of Connecticut} \\
    USA \\
    tingting.yu@uconn.edu}
}
\IEEEpubid{\makebox[\columnwidth]{\textit{MSR 2025, Ottawa, Canada, April 28-29, 2025} \hfill}%
\hspace{\columnsep}\makebox[\columnwidth]{ }}

\maketitle
\begin{abstract}
Automated bug reproduction is a challenging task, with existing tools typically relying on textual steps-to-reproduce, videos, or crash logs in bug reports as input. However, images provided in bug reports have been overlooked.
To address this gap, this paper presents an empirical study investigating the necessity of including images as part of the input in automated bug reproduction. 
We examined the characteristics and patterns of images in bug reports, focusing on (1) the distribution and types of images (e.g., UI screenshots), (2) documentation patterns associated with images (e.g., accompanying text, annotations), and (3) the functional roles they served, particularly their contribution to reproducing bugs. 
Furthermore, we analyzed the impact of images on the performance of existing tools, identifying the reasons behind their influence and the ways in which they can be leveraged to improve bug reproduction.
Our findings reveal several key insights that demonstrate the importance of images in supporting automated bug reproduction.
Specifically, we identified six distinct functional roles that images serve in bug reports, each exhibiting unique patterns and specific contributions to the bug reproduction process. 
This study offers new insights into tool advancement and suggests promising directions for future research.

\end{abstract}


\section{Introduction}
\label{sec:intro}


In the mobile app marketplace, debugging and problem resolution are critical. Research shows that 88\% of app users are likely to abandon an app if they encounter persistent issues, emphasizing the importance of prompt resolution to retain users\cite{applause}. One significant challenge developers face is reproducing bugs reported by users, who often provide insufficient information, such as the exact sequence of their interactions\cite{johnson2022empirical, moran2015auto, bettenburg2008makes, ambriola1997processing}. To tackle this issue, the software engineering community is increasingly focused on automating the bug reproduction process~\cite{fazzini2018automatically, zhao2019recdroid, zhang2024mobile, zhang2023automatically, feng2024prompting, wang2024feedback, huang2023context, huang2024crashtranslator, feng2022gifdroid}.

Bug reproduction relies on the information provided in bug reports to replicate reported issues. 
From a functional perspective, a bug report typically includes essential components such as Steps to Reproduce (S2R), Expected Behavior (EB), and Observed Behavior (OB).  
In terms of media types, it may feature text descriptions, images, and videos/GIFs. 
In this paper, \textit{images} refer to UI page \emph{screenshots} or any \emph{static visual information} provided by the user, which may be beneficial for bug reproduction process.

Table~\ref{table:toolsummary} summarizes the recent state-of-the-art automated bug reproduction tools, highlighting the specific bug report information they utilize and the main techniques employed in their approaches. 
Textual Steps to Reproduce (S2Rs) stand out as the most commonly used information employed by the majority of the works~\cite{fazzini2018automatically, zhao2019recdroid, zhang2024mobile, zhang2023automatically, feng2024prompting}, as they provide the most straightforward instructions needed to trigger the bug. However, other efforts~\cite{wang2024feedback, feng2022gifdroid, huang2024crashtranslator} have shown that information beyond textual S2Rs is critical for bug reproduction, yielding promising results. For instance, Feng et al.~\cite{feng2022gifdroid,feng2023read,feng2022gifdroid1} employ videos and GIFs in bug reports for bug reproduction, while Huang et al.~\cite{huang2024crashtranslator} utilize error log. Similarly, Wang et al.~\cite{huang2023context} leverages the entire textual bug report, including the title and other elements, demonstrating that each piece of information in bug reports can potentially supplement the S2Rs and improve the performance of automated bug reproduction.

\begin{table}[t]
\vspace{10pt}
\centering
\begin{threeparttable}
\caption{\textbf{Summary of Recent Android Bug Reproduction Works}}
\label{table:toolsummary}
\scriptsize
\begin{tabular}{|l||r|c|p{0.8cm}|p{0.825cm}|}
\hline
\textbf{Tool Name} & \textbf{Venue} & \textbf{Input}  & \textbf{Open Source} & \textbf{Dataset Size} \\ \hline \hline

\textbf{Yakusu}~\cite{fazzini2018automatically}   &  ISSTA' 18  & S2Rs        & $\checkmark$ & 60 \\ \hline
\textcolor{red}{$\star$} \textbf{ReCDroid}~\cite{zhao2019recdroid}    &  ICSE' 19  &    S2Rs       &   $\checkmark$    & 51 \\ \hline
\textbf{GIFdroid}~\cite{feng2022gifdroid}&  ICSE' 22  &    GIF \textbackslash Videos      &   $\checkmark$ &   61 \\ \hline
\textcolor{red}{$\star$} \textbf{ReproBot}~\cite{zhang2023automatically}   &  ISSTA' 23  &  S2Rs         & $\checkmark$ & 77  \\ \hline
\textbf{ScopeDroid}~\cite{huang2023context}    & ICSE' 23   &   S2Rs    &  $\times$   &  102   \\ \hline
\textcolor{red}{$\star$} \textbf{AdbGPT}~\cite{feng2024prompting}   &  ICSE' 24  &    S2Rs         & $\checkmark$ &  48   \\ \hline
\textbf{CrashTranslator}~\cite{huang2024crashtranslator}   & ICSE' 24   &  Error Log       & $\checkmark$   & 75  \\ \hline
\textbf{Roam}~\cite{zhang2024mobile}   & FSE' 24   &   S2Rs      & $\checkmark$ &   72    \\ \hline
\textcolor{red}{$\star$} \textbf{ReBL}~\cite{wang2024feedback}   & ISSTA' 24   &   Whole bug report        & $\checkmark$ &  96  \\ \hline
\end{tabular}
\begin{tablenotes}
\footnotesize
\item[1.] \textcolor{red}{$\star$} indicates the tool is selected as baseline for this study.
\end{tablenotes}
\end{threeparttable}
\vspace{-10pt}
\end{table}

However, none of these approaches consider the images provided in bug reports, and the potential for leveraging these images to assist in automated bug reproduction remains unexplored, even though many bug reports include images. For instance, RegDroid~\cite{xiong2023empirical}, which is the most comprehensive dataset of Android functional bugs, features images in 41.35\% (165/399) of its bug reports. 
Similarly, AndroR2~\cite{johnson2022empirical, wendland2021andror2}, the widely-studied dataset for Android bug reports, contains images in 23.33\% (42/180) of its reports. 
%
%
To address this gap, we conducted an empirical study to analyze images in bug reports. Our study is guided by the following primary research questions:



\noindent
\textbf{RQ1: How do the number and types of images vary across bug reports?}
This research question addresses both quantitative (e.g., single-image vs. multiple-image reports) and categorical (e.g., UI screenshots vs. non-UI screenshots) aspects of image usage in bug reports. 
Understanding these patterns is essential for assessing how images contribute to bug reports. This foundation can inform the design of automated bug reproduction tools by identifying image patterns that are likely to be effective in automation or pinpointing specific types of images with the greatest potential to enhance automation processes.



%

\begin{figure*}
    \centering
    \includegraphics[width=0.98\linewidth, height=5cm]{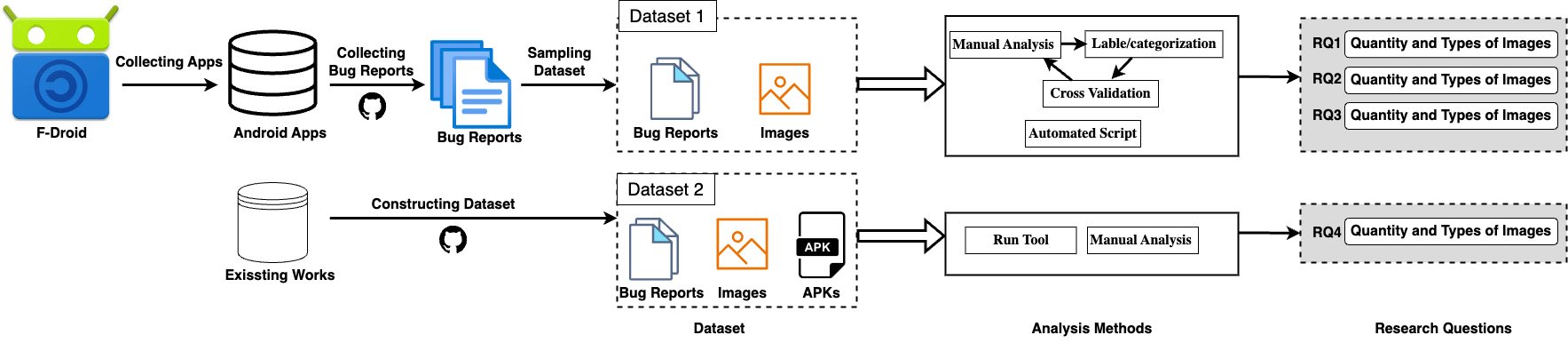}
    \caption{Overview of our study}
    \label{fig:ovwerview}
\end{figure*}

\noindent
\textbf{RQ2: What are the functional roles of images in bug reports?} 
Textual information in bug reports is typically organized into functional categories, such as Steps-to-Reproduce (S2Rs) and Observed Behavior (OB)~\cite{chaparroAssessingQualitySteps2019,chaparroDetectingMissingInformation2017},
each emphasizing different key information and requiring distinct strategies to effectively utilize the information for bug reproduction. For instance, textual S2Rs incorporate specific action verbs (e.g., “click”) and nouns related to UI elements (e.g., “menu”), which have influenced the creation of various tools~\cite{zhao2019recdroid, feng2024prompting, zhang2023automatically}. These tools utilize S2Rs through a phase known as S2R Entity Extraction. During this phase, crucial entities are extracted from S2Rs, identifying important action-noun pairs to facilitate automated bug reproduction.
Similarly, when using crash logs, identifying key information like the activity name where the bug occurred and the specific error exception is crucial for accurately reproducing the bug.
Thus, this research question aims to examine the functional roles that images play in bug reports, identifying patterns within these roles and analyzing the specific information emphasized for each role. This analysis can inspire future research by providing a foundation for leveraging images more effectively: it clarifies how to differentiate between image roles and, for each role, highlights the essential information that should be prioritized or extracted to assist in bug reproduction.

\noindent
\textbf{RQ3: Are images in bug reports documented?} 
While a plain image can convey information, supporting text is essential for a complete understanding, especially when images may serve various functional roles (as explored in RQ2). Correctly identifying these roles for reproduction requires adequate documentation.
Therefore, this research question investigates the extent to which images in bug reports are documented. If documentation is present, it examines its form—such as annotations or accompanying text. When documentation is absent, we need to consider alternative ways to interpret and utilize the image effectively to understand its purpose and relevance.
%
%

\noindent
\textbf{RQ4: What types of images have the most significant impact on the effectiveness of automated bug reproduction?} 
This research question explores the role of images in automated bug reproduction by distinguishing between image types that contribute positively to the process, those with negligible impact, and those that may even hinder effectiveness. We aim to understand why certain images aid or obstruct automation, and to uncover opportunities to leverage these insights—whether to refine existing tools or to design new ones that make more strategic use of image.


Our study highlights the importance of images in automated bug reproduction and presents several key findings regarding the effective use of different image types in bug reports.
In summary, this paper makes the following contributions:

\begin{itemize}

    \item This study represents the first systematic study on the images within bug reports. It investigates the frequency of image inclusion, the functional roles these images serve, how they are documented, and evaluates their impact on the performance of current automated bug reproduction tools.
    
    \item Several key findings offer valuable insights and directions for future research on leveraging images in bug reports to enhance automated bug reproduction and new research direction.

    
    \item We have made the replication package available~\cite{replication}.
\end{itemize}


\section{Methodologies}

This section presents our methodology for dataset collection and analysis. An overview of our study is shown in ~\ref{fig:ovwerview}.

\subsection{Dataset Construction} 
\label{dataconstruction}
We constructed two datasets to address the four research questions. 
The construction of each dataset follows different methodologies suited to their respective objectives. Dataset\textsubscript{1}, designed to answer RQ1 through RQ3, focuses on gathering a wide range of real-world bug reports, including those with images, to analyze their characteristics and variety. This dataset does not require verification of whether the bugs are still reproducible, as its primary purpose is to provide a broad overview of images' characteristics in bug reports rather than current reproducibility verification.
In contrast, Dataset\textsubscript{2}, which is designed to answer RQ4, focuses on conducting experiments using existing tools to reproduce the bug reports. Therefore, this dataset requires manual verification of the bugs to ensure they are reproducible%

\noindent
\textbf{Dataset\textsubscript{1}.}
We adopted established practices~\cite{xiong2023empirical, wendland2021andror2, johnson2022empirical, wang2023empirical} for gathering real-world bug reports to ensure our analysis is based on authentic and practical scenarios.
We initially collected 645 real-world Android apps from F-Droid~\cite{fdroid}, a platform known for its extensive collection of open-source Android apps, most of which are hosted on GitHub.
Using the GitHub REST API~\cite{githubapi}, we crawled 257,140 issues from the GitHub repositories of these 645 apps. To identify bug reports, we applied the following filtering criteria: (1) the issue must include at least one label containing the keyword "bug". (2) to focus exclusively on mobile app-related bug reports, any issue containing the terms "windows", "linux", "desktop", or "tv" (case-insensitive) was excluded from our analysis.
As a result, we obtained a total of 50,988 Android bug reports, of which 7988 (15.67\%) include images. 
To ensure a representative yet manageable dataset for answering our research questions (RQs), we applied standard statistical sampling techniques with a 95\% confidence level and a 5\% margin of error~\cite{illowsky2013introductory}. From the total of 7,988 bug reports containing images in our initial pool of 597 apps, we randomly sampled 367 reports. This finalized Dataset\textsubscript{1}, which is used for answer RQ1, RQ2, and RQ3. 

\noindent
\textbf{Dataset\textsubscript{2.}}
We constructed the second dataset to evaluate the performance of existing tools for handling image-containing bug reports (RQ4). To create this dataset, We followed the dataset construction methods used by established bug reproduction tools~\cite{wang2024feedback, zhang2023automatically, zhang2024mobile, feng2024prompting}, selecting bug reports from a well-known dataset to ensure their representativeness. Additionally, apps in this dataset are well-established within the field of Android bug reproduction. By focusing on these well-studied apps in automated bug reproduction, we leveraged existing knowledge and resources to reduce the time required for manually reproducing
we extended our collection beyond the available reports by examining GitHub repositories to identify more image-containing bug reports.
We applied similar criteria for examining bug reports: (1) they must have accessible APK files, and (2) they must still be reproducible. Additionally, we included one criterion specific to our study: each bug report must contain images, aligning with our focus on image-containing bug reports. As a result, we collected 42 bug reports that include at least one image as Dataset\textsubscript{2} to answer RQ4.


\begin{tcolorbox}[colback=blue!5, colframe=black, boxrule=0.5pt]
\textbf{Finding 1:} The dataset construction process shows that 15.67\% of Android bug reports contain images, emphasizing their frequent inclusion and the need to explore their potential role in automated bug reproduction.
\end{tcolorbox}

\subsection{Analysis Methods.}
To address the research questions, a combination of automated scripts and manual analyses was employed to ensure reliability and validity. 
To understand the prevalence of images in bug reports (RQ1), an automated script scanned each report to identify and count images, followed by a manual classification of images as either UI screenshots or other types of static visualizations. 
To investigate the functional roles of images in bug reports, the authors conducted a multi-round classification process, independently analyzing images to categorize their functions and assess documentation relevance within the bug reports. After each round, they discussed and refined classification categories, adding new ones as needed to reach a consensus.
To analyze the impact of images on the effectiveness of existing bug reproduction tools (RQ3), we tested reports containing images using automated reproduction tools, examining their success rates when image data was omitted and identifying limitations in image handling.
To validate our findings, we used two third-party datasets: AndroR2\cite{wendland2021andror2, johnson2022empirical} and RegDroid\cite{xiong2023empirical}
to ensure that our results were consistent and generalizable.

\subsection{Selecting Bug Report Reproduction Tools} 
\label{tools}

Table \ref{table:toolsummary} summarizes ten recent state-of-the-art studies on automated bug report reproduction, none of which have considered images.  We selected tools that can accept free-text descriptions of bug reports as input. Other tools, such as CrashTranslator \cite{huang2024crashtranslator}, which relies on logcat, GifDroid \cite{feng2022gifdroid}, which utilizes video input, and Roam\cite{zhang2024mobile}, which requires JSON-formatted input, are beyond the scope of our study. 
Our study aims to explore how images within textual descriptions can enhance automated bug reproduction. 
In contrast, logcat outputs and videos represent separate entities within the bug report structure. 
Yakusu\cite{fazzini2018automatically} was excluded due to its limited relevance compared to more recent and advanced tools that have demonstrated superior performance. Additionally, DriodScope\cite{huang2023context} was not included because of its closed-source nature. Consequently, we selected four bug report reproduction tools \cite{zhao2019recdroid, zhang2023automatically, feng2024prompting, wang2024feedback} for our analysis
in RQ3. 

Specifically,
\textbf{ReCDroid}~\cite{zhao2019recdroid} is an automated bug reproduction tool that follows a two-phase approach: S2R entity extraction and S2R entity matching. It combines natural language processing (NLP) with heuristic grammar patterns for entity extraction and employs a guided depth-first search to dynamically explore the application, aiming to match S2R entities with GUI elements to eventually reproduce event sequences that trigger the reported bug. ReCDroid was one of the earliest works in this area and is a popular baseline for evaluating new testing techniques.
\textbf{ReproBot}~\cite{zhang2023automatically} also uses NLP but combines it with reinforcement learning (RL) techniques like Q-learning to optimize the search of the UI actions for crash reproduction. The Q-learning algorithm helps ReproBot to account for missing steps and identifies UI actions that overall match the S2Rs. 
\textbf{AdbGPT}~\cite{feng2024prompting} is the first work that uses Large Language Models (LLMs) through prompt engineering and chain-of-thought reasoning, enabling accurate S2R extraction and replay without extensive training data.
\textbf{ReBL}~\cite{wang2024feedback} bypasses the traditional two-phase approach and the use of S2R entities. Instead, it leverages the entire textual bug report and employs LLMs to guide the reproduction process in a feedback-driven manner. This approach is more flexible and context-aware than the traditional step-by-step entity matching method. It is the first tool capable of reproducing bugs using entire bug reports without relying on S2R entities. 

\section{Empirical Study}

\subsection{Quantity and Types of Images (RQ1)}
This research question examines patterns in the quantity and types of images used in bug reports, offering insights into their distribution and usage. We analyze the average number of images per report, categorize them by type (e.g., UI screenshots vs. other images), and identify how many reports contain single or multiple images. This analysis provides a foundation for further studies to explore the potential of using image usage in bug reports for automated bug production.

In analyzing the 367 bug reports in Dataset$_1$, 70.77\% of reports contained exactly one image, while 29.23\% included multiple images, with the number of images per report ranging from 1 to 7. On average, there were 1.48 images per bug report. Of these, 95.18\% were UI screenshots, while a smaller portion (4.82\%) of these images included pictures of code snippets, crash logs, or photos of the UI screen taken by another device.

 


\begin{tcolorbox}[colback=blue!5, colframe=black, boxrule=0.5pt]
\textbf{Finding 2:} 70.77\% of reports included one image, while 29.23\% had multiple images (ranging from 1 to 7 per report). Of all images, 95.18\% were UI screenshots, and 4.82\% were non-UI (e.g., code snippets, error logs).

\end{tcolorbox}

Since the majority of reports contain a single image, it is crucial to understand the functional role of this image in single-image reports. What kind of information does it provide, and could it effectively be leveraged in automated bug reproduction? Conversely, in reports with multiple images, it is important to consider whether these images serve the same functional role or convey different aspects of the bug, as well as to examine the logical sequence between images. With these questions in mind, we will further explore the functional roles of images in RQ2.

\begin{figure*}[t]
    \centering
    \vspace{10pt}
     \includegraphics[scale=0.3]{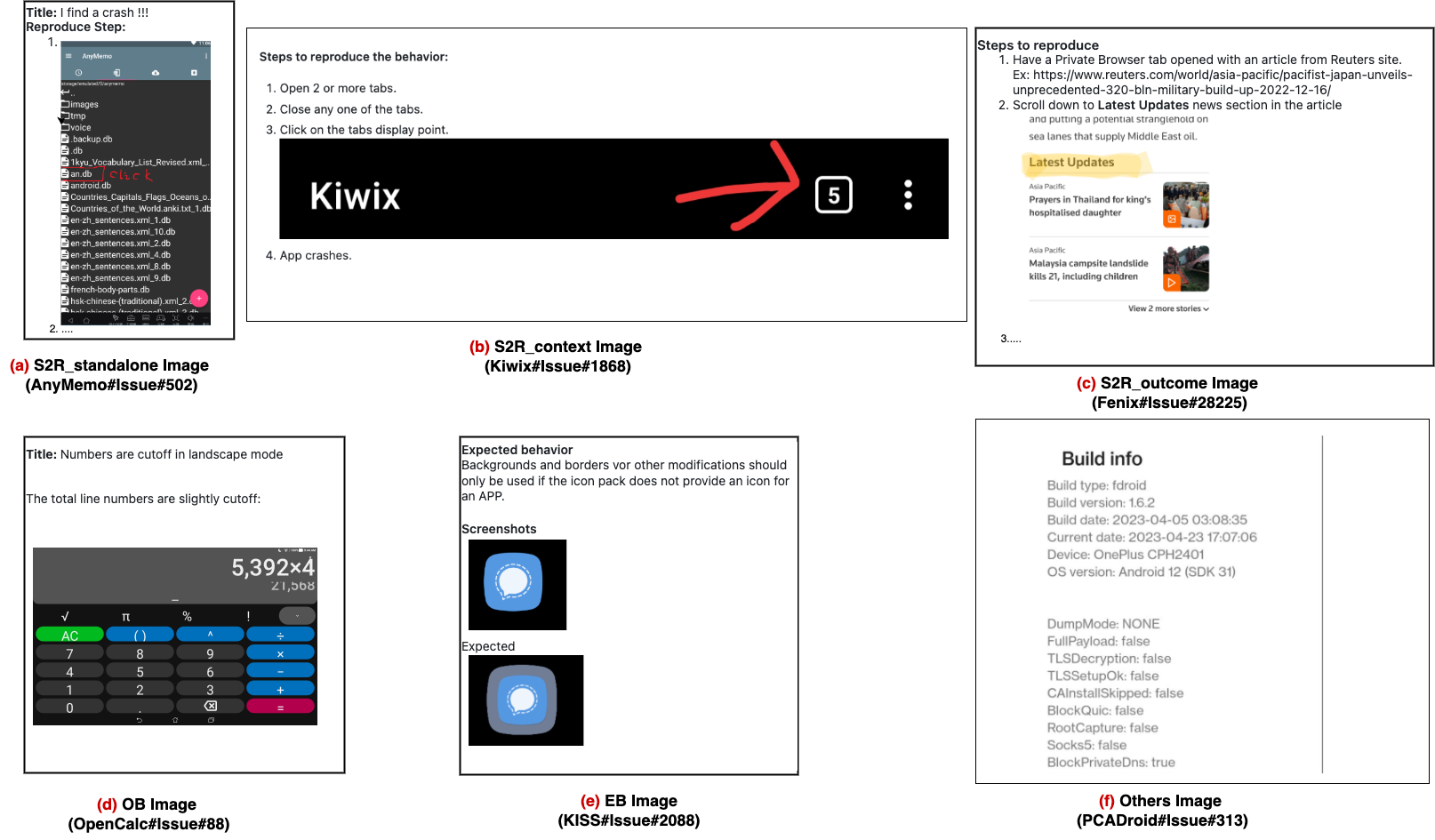}
    \caption{Examples of Images in Different Functional Roles}
    \label{fig:roles}
\end{figure*}

\subsection{Functional Roles of Images (RQ2)}
\label{rq2}


%

%
Bug reports typically consist of several key components that describe an issue, each serving a specific purpose: Steps to Reproduce (S2Rs), Observed Behavior (OB), and Expected Behavior (EB)~\cite{chaparroAssessingQualitySteps2019,chaparroDetectingMissingInformation2017}. 
Following this structure, images in bug reports can be classified into four main categories: \textit{S2R}, \textit{OB}, \textit{EB}, and \textit{Others}. The \textit{Others} category includes images that do not fit into the first three categories (e.g., screenshots of code). 
Additionally, we found that images in \textit{S2R} category could be further categorized into three distinct types based on their characteristics: \textit{S2R$_{standalone}$}, \textit{S2R$_{context}$}, and \textit{S2R$_{outcome}$}.  
Just like textual descriptions in bug reports, bug reproduction tools that can correctly classify images into their respective functional roles are essential for successful bug reproduction.

\subsubsection{
\textbf{\textit{S2R$_{standalone}$ image}}} 
An \textit{S2R\_standalone image} is a self-contained visual representation of an S2R that requires no accompanying text. 
These images are typically annotated to emphasize target user interface (UI) elements and the actions to be performed. 
For instance, in Fig. ~\ref{fig:roles}-(a), the first S2R is depicted entirely through an annotated image: the target element "an.db" is circled in red, and the action "click" is clearly indicated next to it. This effectively communicates the reproduction step "click on an.db." Images that fully capture an entire step-to-reproduce (S2R) without any accompanying text are important for effective bug reproduction. 

Neglecting such images could lead to missing steps, a major factor impacting the performance of automated bug reproduction processes. 

\subsubsection{
\textbf{\textit{S2R$_{context}$ image}}} 
An \textit{S2R$_{context}$ image}  complements textual S2R instructions by providing additional visual context. 
Typically, it accompanies text in a pattern, such as "textual S2R + image". 
The textual S2R offers the primary guidance for reproducing the step, while the image adds crucial visual cues, forming a complete S2R together. 
These images often highlight target UI elements or required text inputs (e.g. Figure~\ref{fig:roles}-(b)).
Users tend to include images to depict target elements that are difficult to describe, especially when the element is uncommon or lacks a clear label—such as a menu icon or a "more options" button. 
When images are used to show required text input, it often indicates that the input is complex or specially formatted, not just simple text that can be easily typed in the bug report.
For example, in the bug report shown in Figure~\ref{fig:roles}-(b)), the third S2R states, “Click on the tabs display point,” followed by an annotated image that indicates which UI element is the "tabs display point," helping the developer quickly and accurately locate the target element. 

Such images reduce ambiguity in S2R instructions by providing visual cues of the target, enhancing clarity and understanding. This addresses another main challenge in existing work, where the bug report written by the user might be ambiguous and not clear due to their writing habits and lack of technical background.

\subsubsection{
\textbf{\textit{S2R$_{outcome}$ image}}} 
This type of image captures the state of the UI after a S2R has been replayed.  Unlike other S2R images that either represent S2Rs or are used to complement S2Rs, \textit{S2R$_{outcome}$ images} focus solely on displaying the results following the execution of the S2R.
In manual bug reproduction, an \textit{S2R$_{outcome}$ image} is primarily used to verify whether an S2R has been successfully replayed by comparing the image with the actual UI page. This visual confirmation allows developers to ensure that the action produced the expected outcome, thereby validating the accuracy of the reproduction step.

However, in automated bug reproduction, existing tools typically do not perform any verification of S2R replaying.  Instead, these tools rely solely on the presence of the target of the next S2R to advance through the reproduction steps. For example, given two steps, $S1$ and $S2$, after performing $S1$, the tool searches for the target element of the subsequent S2R, $S2$, within the current UI state. If the element for $S2$ is not found, the tool may backtrack to a previous UI state to attempt replaying $S1$ again. If the target element is found, the tool proceeds to replay $S2$ and continues with the subsequent steps.
This approach assumes that the mere presence of next target elements is sufficient to determine the success of each step. Consequently, the utility and necessity of \textit{S2R$_{outcome}$ images}  in automated bug reproduction remain unexplored. While these images undeniably offer valuable visual confirmation in manual processes, their integration into automated systems has not been considered. Determining whether incorporating \textit{S2R$_{outcome}$ images} would enhance automated bug reproduction requires experimental validation and further investigation. 
\begin{tcolorbox}[colback=blue!5, colframe=black, boxrule=0.5pt]
\textbf{Finding 3:} 
S2R images fall into three subcategories—context, standalone, and outcome—based on how they convey information. S2R$_{context}$ and S2R$_{standalone}$ images reduce ambiguity by clarifying reproduction steps and providing visual cues to support textual guidance. 
S2R$_{outcome}$ images provide essential visual confirmation in bug reproduction but are currently overlooked by existing tools.
\end{tcolorbox}

\subsubsection{
\textbf{\textit{OB image}}} 
\textit{OB image} visually illustrates the reported bug symptom,  validating the existence of the bug. This saves developers from spending time questioning the existence of the bug when they are unable to reproduce it.
From the perspective of bug reproduction, OB images serve dual purposes: visually illustrating bug symptoms and aiding the step replay. OB images clearly convey the bug symptoms, helping developers comprehend the issue and recognize when the bug is triggered during manual reproduction. Moreover, OB images provide additional context and visual cues to guide navigation to the target page where bugs occur. 

For example, the images in Figure~\ref{fig:roles}-(d) shows OB images, illustrating the non-crash bug symptoms. 
In  Figure~\ref{fig:roles}-(d), the bug report shows the OB image with short description but not   S2Rs for reproducing the bug. Nevertheless, the image provides useful context: aside from displaying the overlap bug symptom, the image reveals that this issue occurred on the calculator page. The digits shown on the screen imply that to replicate the overlap, it might be necessary to input an equation in the calculator. Thus, beyond highlighting the bug symptom, \textit{OB images} offers clues about the specific page where the overlap took place.

To be clear, although \textit{OB images} and \textit{S2R$_{outcome}$ images} share similarities, as both capture the UI results of specific steps in the reproduction process, the key distinction lies in whether the image displays the reported bug symptom. An \textit{OB image} is essentially a subset of \textit{S2R$_{outcome}$ images}, specifically capturing the result of the final S2R where the bug symptom is visible. In contrast, \textit{S2R\_outcome images} may depict the result of any step along the reproduction process and do not show the bug symptom. Therefore, if an image is an \textit{S2R\_result images} of the last S2R, it is classified as an \textit{OB images} rather than an \textit{S2R\_outcome images}.

\subsubsection{
\textbf{\textit{EB image}}} 
An \textit{EB image}  provides a visual reference of the expected or correct behavior,  often captured from a previous version of the application. In our dataset, we found that bugs related to issues like padding, text alignment, and color settings (including dark and day modes) frequently include EB images to help developers visually compare and identify the discrepancies. Additionally, EB images often appear alongside OB images for direct comparison, as illustrated in Figure~\ref{fig:roles}-(e).

Despite this benefit, existing automated tools ignore not only EB images but also any textual information related to EB. This is because EB information serves as a reference for how the application should behave, rather than explicitly showing a failure. This implicit nature makes it challenging for automated tools to utilize, unlike S2Rs (which provide direct instructions for reproducing the bug) or OB images (which display the exact symptoms of the issue). 
However, we should not overlook the potential of leveraging EB images to enhance automated bug reproduction. One possible integration is to use EB images in conjunction with OB images for automated oracle checking. By comparing EB images (representing correct behavior) with OB images (displaying bug symptoms) and the actual UI page, automated tools can more effectively identify discrepancies. Another approach is to view the EB image as the inverse of the OB image, suggesting that the EB image also describes the same UI page. This perspective allows EB images to implicitly provide information about the UI page (e.g., the target state) where the bug occurs. 

\subsubsection{
\textbf{\textit{Others image}}} 
These images can be classified as follows: (i) non-UI images, such as screenshots of source code or error logs; (ii) combined screenshots, which are single images created by merging multiple screenshots, making them challenging to interpret due to the diverse information they contain and the lack of clear context, and the difficulty of separating them; and (iii) UI images that are not directly related to OB, EB, or S2R, such as a screenshot of the settings screen that simply shows the application's configuration (e.g., Figure~\ref{fig:roles}-(f)).
Although not directly illustrating the bug, these images (e.g., logs) can be valuable by revealing settings or conditions impacting the application’s behavior.


\begin{table}[h]
\centering
\caption{Distribution of Image Roles Across Single-Image and Multiple-Image Bug Reports}
\label{tab:rq2}
\small
\begin{tabular}{|l|c|c|c|}
\hline
 \rowcolor{gray!45}     Image Roles       & Sgl-Img BR & Multi-Img BR &  \textbf{Overall}\\ \hline\hline
 
\textit{S2R\_standalone}   & 0  & 2.60\% &  0.82\%\\ \hline
\textit{S2R\_context}   & 1.79\%  &  20.78\%  &   7.76\%\\ \hline
\textit{S2R\_outcome}   & 1.19\%  & 10.39\% &  4.08\% \\ \hline
\textit{OB}  &  93.45\% &  90.91\% & 92.65\% \\ \hline
\textit{EB}  &  2.38\%  &  46.75\% & 16.33\% \\ \hline
\textit{Others}   & 1.19\%  & 20.78\% & 7.35\% \\ \hline

\end{tabular}
\vspace{-10pt}
\end{table}

Table~\ref{tab:rq2} shows distribution of image roles
across image bug reports. The percentages refer to the proportion of bug reports containing images in each functional role, rather than the percentage of each image type over the total number of images.
In the Overall column, the majority of bug reports(92.65\%) contain at least one image serving the OB role, followed by EB images and S2R images. This distribution emphasizes the popularity of using images to visually present observed and expected outcomes, likely due to the variety of non-crash symptoms and expected behaviors that are more effectively conveyed through visuals rather than textual descriptions. Images provide a clearer representation of the issues and the desired results, especially in cases where complex visual elements are involved, making them a valuable resource in the bug reporting and reproduction process.

We further examine the functional roles of images in bug reports by differentiating between those that contain a single image and those with multiple images.  We hypothesize that this separation will aid in classifying images into their respective functional roles more effectively. For instance, if most bug reports with a single image predominantly feature OB images, it would be simpler to categorize these images accordingly. As the result, the last two columns of Table~\ref{tab:rq2} suggest the following:

First, \textit{OB} are dominant in both single-image and multiple-image bug reports, with 93.45\% of single-image reports and 90.91\% of multiple-image reports containing at least one OB image. This highlights the popularity of using images to illustrate observed behavior, underscoring their value in effectively communicating the symptoms of a bug. Additionally, we can infer that when a bug report includes only one image, it is most likely to be an \textit{OB} image, suggesting that capturing the observed behavior is often the primary purpose of single-image bug reports.

Second, apart from \textit{OB} images, the proportion of all other roles increases significantly in multiple-image reports. This can be attributed to two main reasons: (1) when a bug report includes more than one image, there is greater diversity in the roles these images can fulfill, and (2) in multiple-image bug reports, each image can serve a different role, leading to overlapping classifications. Consequently, the cumulative percentage for each role exceeds 100\% as individual reports can contain multiple images fulfilling multiple functional roles. This highlights how multiple-image bug reports are often more comprehensive, capturing various aspects of the bug, such as observed behavior, expected behavior, and contextual information, which collectively enhance the clarity and reproducibility of the bug report.


\begin{tcolorbox}[colback=blue!5, colframe=black, boxrule=0.5pt]
\textbf{Finding 4:} 
First, the majority of bug reports (92.65\%) contain at least one OB image. This trend is especially strong in single-image reports, where 93.45\% are OB images, suggesting that when only one image is included, it is most likely to represent the observed issue. 
Second, multiple-image reports contain diverse images serving different roles, which implies that if future work aims to leverage these images to facilitate automated bug reproduction, it will be crucial to understand how to identify and differentiate these roles effectively.
\end{tcolorbox}

%




\subsection{Documentation of Images (RQ3)}
During the classification of the functional roles of images (RQ2), it is evident that static images provide limited information on their own. To accurately differentiate the roles of these images, it is necessary to consider the accompanying textual information in the bug report. This additional context is important for interpreting the intent and relevance of the images. Therefore, documentation becomes a key aspect. We studies the documentation of images—specifically, those that include annotations or are accompanied by explanatory text. Such documentation helps indicate the purpose or context of the image, enhancing its utility in understanding and reproducing the bug.
An image was considered documented following this criteria: we first determined whether an image contained Explanatory Text. If it did, we classified it under this category. If an image lacked explanatory text, we then assessed whether it was placed under a section with descriptive subsection titles. If an image met neither of these criteria, it was categorized as a plain image.
\begin{itemize}[left=0.1cm]
    \item \textbf{Explanatory Text (38.06\%).} Descriptive text accompanying the image that explains what the image shows or how it relates to the issues being reported.

    \item \textbf{Descriptive Subsection Titles (5.17\%).} Clear and informative headings above the image that indicate what the image is illustrating. Titles like "Observed Behavior" are helpful, whereas generic titles like "Screenshot" do not provide meaningful context. (e.g., plain images but placed under a subsection with a descriptive title)

\end{itemize}

\noindent
Of the images in bug reports, 43.23\% were identified as documented images, while 56.77\% were categorized as plain images. Plain images typically occur in two scenarios: (1) multiple consecutive images are placed together without accompanying text to clarify their differences or relationships, and (2) many bug reports use a template with sections labeled “Screenshots” or “Other.” Although these sections are well-intentioned, users often upload screenshots without additional explanation, making it difficult to interpret the images and challenging to identify which parts of the text they are intended to support.

\begin{tcolorbox}[colback=blue!5, colframe=black, boxrule=0.5pt]
\textbf{Finding 5:}
Documented images (43.23\%) with explanatory text or descriptive titles enhance clarity, while plain images (56.77\%). If one intends to use images in automated bug reproduction, this lack of clear context complicates the classification process, making it challenging to accurately interpret each image's functional role.
\end{tcolorbox}

\begin{table*}[h]
\caption{Experimental Results for RQ4.}
\label{tab:rq4}
\centering
\begin{threeparttable}
\small
\begin{tabular}{|>{\columncolor{gray!30}\bfseries}c|l|c|c|c|c|>{\columncolor{gray!30}\bfseries}c|l|c|c|c|c|}
\hline
\rowcolor{gray!30} ID & Bug Report  & RD &  RB & AG   & BL  & ID & Bug Report  & RD &  RB & AG  & BL  \\ 
\hline
1 & AFM-1657 & $\times$  & $\times$ & $\times$ & $\times$ & 22 & AFM-1794  & $\checkmark$ & $\checkmark$ &$\checkmark$  & $\checkmark$ \\ 
2 & AFM-1795  & $\checkmark$  & $\checkmark$ & $\checkmark$ & $\checkmark$ & 23 & AndOTP-827  &  $\times$ &  $\times$ &  $\times$  & $\times$ \\ 
3 & AFM-1796  & $\times$  & $\times$ & $\checkmark$ & $\checkmark$ & 24 & AndrOBD-243  & $\checkmark$ & $\checkmark$ &$\checkmark$  & $\checkmark$ \\ 
4 & AFM-2477  & $\times$  & $\times$ & $\times$ & $\times$ & 25 & Android-1248 &  $\times$ &  $\times$ &  $\times$  & $\times$ \\ 
5 & andOTP-551  & $\times$  & $\times$ & $\times$ & $\times$ & 26 & AnyMemo-500 &  $\times$ &  $\times$ &  $\times$  & $\times$ \\ 
6 & andOTP-827  & $\times$  & $\times$ & $\times$ & $\times$ & 27 & Fenix-27725  &  $\times$ &  $\times$ &  $\times$  & $\times$ \\ 
7 & Anki-13919  & $\times$  & $\times$ & $\times$ & $\times$ & 28 & Fenix-27987 &  $\times$ &  $\times$ &  $\times$  & $\times$ \\ 
8 & Anki-14609  & $\times$  & $\times$ & $\times$ & $\times$ & 29 & Fenix-28086  &  $\times$ &  $\times$ &  $\times$  & $\times$ \\ 
9 & Anki-16325  & $\times$ & $\checkmark$ & $\times$ & $\times$ & 30 & k9-4804  &  $\times$ &  $\times$ &  $\times$  & $\times$ \\ 
10 & AnkiDroid-6228   & $\times$ & $\times$ & $\times$ & $\times$ & 31 & Markor-1815 &  $\times$ &  $\times$ &  $\checkmark$  & $\checkmark$ \\ 
11 & AnkiDroid-7286   & $\times$ & $\times$ & $\times$ & $\times$ & 32 & Markor-1961  &  $\times$ &  $\times$ &  $\times$  & $\times$ \\ 
12 & AnkiDroid-9005   & $\checkmark$ & $\checkmark$ & $\checkmark$ & $\checkmark$ & 33 & NewPipe-10090  &  $\times$ &  $\times$ &  $\times$  & $\times$ \\ 
13 & AnyMemo-502  &  $\times$ & \textcolor{red}{$\circ$} & \textcolor{red}{$\circ$} & \textcolor{red}{$\circ$}  & 34 & NewPipe-10380 &  $\times$ &  $\times$ &  $\times$ & $\times$ \\ 
14 & Kiwix-1868 & $\times$  & $\times$ & $\times$ & $\times$ & 35 & NewPipe-10646  &  $\times$ &  $\times$ &  $\times$  & $\times$ \\ 
15 & Markor-1020   & $\times$  & $\times$ & $\times$ & $\checkmark$ & 36 & NewPipe-41  &  $\times$ &  $\times$ &  $\times$  & $\times$ \\ 
16 & Markor-1565   & $\times$  & $\times$ & $\times$ & $\times$ & 37 & NewPipe-56  &  $\times$ &  $\times$ &  $\times$  & $\times$ \\ 
17 & Markor-1729   & $\times$  & $\times$ & $\times$ & $\times$ & 38 & News-156  &  $\checkmark$&  $\checkmark$ &  $\checkmark$  & $\checkmark$ \\ 
18 & Markor-550   & $\times$ & $\times$ & $\times$ & $\times$ & 39 & OpenNoteScanner-166  &  $\times$ &  $\times$ &  $\times$  & $\times$ \\ 
19 & Markor-567   & $\times$ & $\times$ & $\times$ & $\times$ & 40 & OpenNoteScanner-39  &  $\times$ &  $\times$ &  $\times$  & $\times$ \\ 
20 & OmniNotes-634   & $\times$ & $\times$ & \textcolor{red}{$\circ$} & \textcolor{red}{$\circ$} & 41 & Aopentasks-898  &  $\checkmark$&  $\checkmark$ &  $\checkmark$  & $\checkmark$ \\ 
21 & Trickytripper-49  & $\times$ & $\checkmark$ & $\checkmark$ & $\checkmark$& 42 & PdfViewer-55  &  $\times$ &  $\times$ &  $\times$  & $\times$ \\ 
\hline
\end{tabular}

\begin{tablenotes}
\item[1] $\checkmark$ Reproduction succeeds without considering the image.
\item[2] $\circ$ Reproduction succeeds only after manually providing the information of the image.
\item[3] $\times$ Reproduction failed in both cases.
\end{tablenotes}

\vspace{-10pt}

\end{threeparttable}
\end{table*}

\subsection{Existing Tools (RQ4)}

In this section, we assess how images in bug reports influence the effectiveness of automated bug reproduction through two experiments. First, we examine whether existing tools, which currently lack the capability to process images as input, can successfully reproduce bugs when images are omitted from the reports. 
Second, we explore whether the functional content of images, if ideally translated into descriptive text, can enhance the accuracy and effectiveness of bug reproduction with existing tools.

\subsubsection{Experiment Design and Setup}
The first experiment uses textual information from bug reports as input to determine whether existing tools could automatically reproduce bug reports.  If the tools successfully reproduced the bugs within a reasonable timeframe without utilizing the images from the bug reports, it would suggest that the images may not be critical, as the textual hints provided are sufficient to guide the existing tools. The second experiment focused on the cases where the first experiment failed. In these cases, we manually turned the information from the images into text descriptions. We divided the images into two groups: S2R and non-S2R. S2R Images provide clues about specific actions, targets, or elements involved in the reproduction steps. We translated these images into precise action-target tuples that existing tools can interpret.  In contrast, non-S2R images, which lack specific actions and targets and cannot be directly converted into action-target tuples, were handled by listing the names of the UI elements on the images. This included details such as the page title (if any) and the various UI elements present in the image. If supplementing the translated information leads to successfully reproducing previously failed cases, it suggests that precisely capturing the image’s intent and extracting key information can effectively leverage images to enhance bug reproduction.

We conducted our experiment on a physical x86 machine running Ubuntu 16.04, equipped with an i7-4790 CPU @ 3.60GHz and 32 GB of memory. We followed the same evaluation settings as in ReCDroid~\cite{zhao2019recdroid}, 
ReproBot~\cite{zhang2023automatically}
AdbGPT~\cite{feng2024prompting}, and ReBL~\cite{wang2024feedback} limiting each experiment to a maximum runtime of one hour. If an experiment exceeds this time limit, it is terminated and marked as a failure. For ReBL~\cite{wang2024feedback}, which includes a summarization mechanism, we maintained the same experimental settings, setting the summarization threshold to three iterations.

\subsubsection{Results} 
Table~\ref{tab:rq4} shows detailed results of the bug reproduction for each bug report by the state-of-the-art tools. 23.8\% of bug reports were successfully reproduced by at least one tool, while 76.2\% were not reproduced by any existing tools. We further analyze the failed cases and summarize the following key observations.


\begin{itemize}[leftmargin=0.3cm]

\item  
\textbf{\emph{ \textit{S2R$_{context}$ image} and \textit{S2R$_{standalone}$ image} play a critical role in filling in missing steps during the bug reproduction process.} (Example: AnyMemo\#502).} As introduced in RQ2 of our empirical study (Section~\ref{fig:roles}) S2R${standalone}$ represents a complete Step-to-Reproduce (S2R) image that visually illustrates specific actions, while an S2R${context}$ image complements the textual S2R by providing additional context. 
Our experimental results validate these findings, showing that omitting these images often leads to incomplete information or overlooked steps, resulting in failed reproduction attempts. 
Additionally, in a second experiment, we attempted to convert the visual information from images into action-target components to enhance the textual reports. This approach led to successful reproduction in two previously failed cases, demonstrating the potential of leveraging S2R images to complete the S2Rs. This proof-of-concept suggests that developing methods to translate S2R images into actionable text could help address these challenges and harness the benefits of S2R images in future automated bug reproduction.

\item 
\textbf{\emph{Images in bug reports are less critical if the textual S2Rs are sufficient.} (Example: Trickytripper\#49).} 
Sufficient information refers to Steps to Reproduce (S2Rs) that are comprehensive and detailed, providing all the necessary target-action information required for accurately reproducing the bug.
In these cases, images become less critical in automated bug reproduction, as existing tools have made significant progress using S2Rs alone, applying various techniques (e.g., reinforcement learning, and prompt engineering) and achieving promising results. This raises an interesting question: can we assess if textual information is sufficient for bug reproduction before running the tool? If so, we could selectively use text-only tools or incorporate images as needed. Beyond this, the classifier could help determine whether to leverage LLM-based~\cite{wang2024feedback, feng2024prompting} or NLP-based~\cite{zhao2019recdroid} tools, depending on whether the textual information is structured and the action-target details are clearly documented, opt for tool using crash log as input if a crash log is available~\cite{huang2024crashtranslator}, or apply video-based reproduction tools~\cite{feng2022gifdroid} when videos are provided.
Therefore, it might be beneficial to build a classifier to assess information sufficiency, enabling more efficient tool selection and improved performance.


\item 
\textbf{\emph{The role of non-S2R images needs to be further determined.}} 
S2R images can be accurately translated manually into a textual format, such as action-target pairs, which existing tools can readily use. This structured format is effective because it aligns well with the current capabilities of automated tools. However, when images convey information beyond simple action-target pairs—such as OB (Observed Behavior) images that illustrate the varying symptoms or EB (Expected Behavior) images that depict the desired outcome—it becomes challenging to convert them accurately into text, and text alone cannot intuitively capture these nuances. A significant challenge in existing works is the bug oracle, which requires a comprehensive understanding of OB information. Therefore, exploring OB and EB images might be a breakthrough in addressing this challenge, as incorporating images can supplement and enrich the textual OB and EB information.

\end{itemize}

\begin{tcolorbox}[colback=blue!5, colframe=black, boxrule=0.5pt]

\textbf{Finding 6:} 
First, S2R${context}$ and S2R${standalone}$ images are essential for automated bug reproduction and can be effectively translated into textual tuples, aligning well with the current capabilities of automated tools.
Second, non-S2R images are more complex because they lack the clear action-target patterns found in S2R images, making them harder to interpret. These images often convey nuanced information, such as observed or expected behaviors, which require further investigation to effectively integrate into automated bug reproduction.
\end{tcolorbox}

\section{Representativeness of Study} 
\label{sec:general}
In this section, we validated the representativeness of our study for RQ2 to ensure that our findings are not limited to the specific characteristics of the initial sample. We performed this validation for RQ2 specifically because it involves classifying images into six distinct roles, a more nuanced task than RQ1 and RQ3, which focus on characteristics like quantity and documentation.
We validated our results by applying the same analysis to two independent third-party datasets 
 AndroR2~\cite{wendland2021andror2, johnson2022empirical} and RegDroid~\cite{xiong2023empirical}.  
 The two datasets were not constructed by the authors of this paper, and the presence of images in bug reports was not one of the criteria in their construction, nor were images in bug reports studied in subsequent analyses.

 AndroR2 contains 23.33\% (42/180) of bug reports with images, though three of these reports either no longer have access to the images or the images are externally hosted. For RegDroid, 41.35\% (165/399) of bug reports include images, but five of these reports lack access to the images. We focused on bug reports with accessible images, which include 36 from AndroR2 and 160 from RegDroid. We applied the same methodologies as used in  RQ2  to classify the functional role of images in the bug reports. The results were then compared to our own dataset to determine if the findings from these third-party datasets align with ours. This analysis helps assess the representativeness of our findings.

\begin{table}[h]
\centering
\caption{Distribution of Image Roles Across Single-Image and Multiple-Image Bug Reports}
\label{tab:genreal}
\small
\begin{tabular}{|l|c|c|c|}
\hline
 \rowcolor{gray!45}            & AndroR2 & RegDroid &  \textbf{Our Dataset}\\ \hline\hline
 
\textit{S2R\_standalone}   & 0  & 0.6\% &  0.8\%\\ \hline
\textit{S2R\_context}   & 9.8\%  &  11.1\%  &   7.8\%\\ \hline
\textit{S2R\_outcome}   & 2.9\%  & 8.3\% &  4.08\% \\ \hline
\textit{OB}  &  93.45\% &  100\% & 92.65\% \\ \hline
\textit{EB}  & 18\%  &  5.6\% & 16.33\% \\ \hline
\textit{Others}   & 6.5\%  & 5.6\% & 7.35\% \\ \hline

\end{tabular}

\end{table}
 Table~\ref{tab:genreal} presents the results of classifying image roles of our dataset and the two third-party datasets. The findings reveal a consistent trend: most bug reports feature OB images, following EB and S2R images. The similar distribution of image roles across all three datasets indicates that our study's findings are generalizable beyond the randomly sampled dataset. By classifying and evaluating images in AndroR2 and RegDroid, we demonstrate that the identified patterns are consistent across different data sources and contexts. This cross-validation confirms the robustness of our findings regarding the utilization of images in bug reports, thereby enhancing the validity and reliability of our conclusions.

\section{Implications and Opportunities}

In this section, we explore the implications of our study, emphasizing the contributions of images in bug reports and how developers and researchers can leverage them to improve automated bug reproduction.

\subsection{Bug Reproduction}
\noindent
\textbf{\emph{Implication 1: Understanding the various types of images and identifying their roles is important.}}
RQ2 findings indicate that different images serve distinct functional roles, each providing unique information, while RQ3 demonstrates that images in multi-image bug reports fulfill varied purposes. This complexity makes it difficult to balance the information presented in both text and images within a bug report. How can we effectively identify the roles of these images and understand their relationship to the accompanying text?

\textbf{\textit{Opportunities:}} A human-like tool could employ a multi-agent system, where each agent is responsible for handling textual information and images according to their distinct functional roles. This collaborative approach enables agents to independently analyze the different roles of images and text before combining their findings to make decisions for accurate bug reproduction.

\noindent
\textbf{\emph{Implication 2: Images can complement S2Rs to minimize the risk of missing steps.}}
When considering images in bug reports for automating bug reproduction, it is essential to at least consider S2R images, just as most existing tools focus on the textual S2R for bug reproduction while overlooking other textual information. S2R images are critical for step replay, especially for S2R$_{context}$ images and S2R$_{standalone}$ images
(Finding 6). 
The context images provide necessary target information or input details, while standalone images represent complete steps on their own. Without these standalone images, it's like creating a missing step, which increases the challenge of bug reproduction. Although some existing tools can bridge missing steps, incorporating images can improve the efficiency of the process.

\textbf{\textit{Opportunities:}} 
To effectively utilize images in bridging the gaps of missing steps in automated bug report reproduction, we can apply Optical Character Recognition (OCR) techniques~\cite{mittal2020text} and heuristic patterns to extract key information from S2R images, such as target elements and actions. 
In addition, we can implement a multimodal approach with reasoning capabilities that generate suggestions based on the available textual S2R and the corresponding S2R image.

\noindent
\textbf{\emph{Implication 3: Images could be a breakthrough in the development of automated reproduction for non-crash bugs.}} Non-crash bug reproduction presents a significant gap in this field, particularly due to the diverse symptoms associated with non-crash bugs~\cite{xiong2023empirical, wang2022detecting, wang2024feedback, baral2024automating}. There are many oracle techniques, each targeted at certain types of bug.~\cite{baral2024automating, su2021owleyes, guo2022ifixdataloss, sun2021setdroid, su2021fully, wang2022detecting, escobar2020empirical, fazzini2017automated, ju2024study}. Existing bug reproduction works predominantly targets crash bugs, largely because of the lack of effective verification capabilities for non-crash bug reports, which makes them more difficult to diagnose and reproduce. Recent studies have only begun to explore the potential of large language models (LLMs) for handling non-crash bug reports reproduction, yet they have not conducted detailed evaluations or analyses. 
As a result, there is an urgent need for tools that can reproduce non-crash bug reports, as well as more general tools for bug symptom verification. Utilizing images could be instrumental in addressing this challenge, especially when the focus shifts to verification rather than merely replaying steps. Our study has shown that many images in bug reports capture non-crash observations and expected behavior, providing valuable visual insights (Findings 4). By analyzing these non-crash images, we can gain a deeper understanding of non-crash behaviors, enabling us to detect non-crash symptoms in the user interface (UI) more effectively, rather than solely identifying non-crash states within the UI.

\textbf{\textit{Opportunities:}} 
To automatically reproduce non-crash bug reports, we can leverage multimodal learning to capture non-crash symptoms. This involves exploring whether a pre-trained model can accurately detect bug symptoms and compare the user interface (UI) with the Observed Behavior (OB) image through prompt engineering, or if fine-tuning the model is necessary to improve performance.


\noindent
\textbf{\emph{Implication 4: Images should be placed in their appropriate sections.}}
RQ3 investigates the documentation of images and highlights a common issue: many images are presented without sufficient explanatory context. Plain images that lack accompanying text or are placed under generic, non-descriptive subsection titles create challenges for users, who must sift through the entire bug report to discern the image's purpose and its relationship to specific text. For instance, a plain image in a generic “Screenshot” section requires readers to infer whether it pertains to Steps to Reproduce (S2Rs), Observed Behavior (OB), or is unrelated. This ambiguity makes it difficult to understand the image's functional role and purpose in the bug report, ultimately hindering its effective use.


\textbf{\textit{Opportunities:}} Methods can be developed to help users place images close to relevant text or sections of the bug report, facilitating a clearer understanding of the visual content in relation to the issue. It may not be a good idea to create a standard template for additional or screenshot sections; however, if a screenshot section is necessary, users should be encouraged to provide explanations for each image. Simple statements like “This is what I observe” or “This relates to Step 1” can offer valuable context. This additional information will assist automated techniques, such as machine learning, in determining the appropriate role of each image.

\section{Threats to Validity}
Our study may suffer from some threats to validity. 
First, although we used a 95\% confidence level, some unique or less common characteristics of bug reports may still be underrepresented, potentially impacting our findings. 
Second, the relatively small number of bug reports in RQ4 may limit the generalizability of our conclusions. This limitation is further reflected in the size of the S2R dataset, which, while imbalanced, mirrors real-world scenarios. The limited number of bug reports containing S2R may impact the generalization of our findings, similar to other empirical studies. However, as this is the first dataset of image-based bug reports, we plan to expand its size as part of our future work.
Third, the manual classification of image roles introduces potential for human error, even with cross-validation, which may impact labeling consistency. To enhance generalizability, we analyzed two third-party datasets, AndroR2 and RegDroid, to classify the functional roles of images within these datasets. The results from both datasets are consistent with our findings.
Fourth, we evaluated two recent tools, AdbGPT and ReBL, in RQ4. Both tools use GPT~\cite{chatgpt} models, which are continually updated, meaning that the results may not remain consistent over time. Additionally, our findings highlight a critical limitation of current tools: they are primarily designed to process textual S2R information, making them less effective at leveraging the rich information within non-S2R images to improve bug reproduction.

%
%


\section{Related Work}

\noindent\textbf{Automated Bug Report Reproduction.}
There are existing approaches that specifically target automatically reproducing Android bug reports as listed in Table~\ref{table:toolsummary} in Section~\ref{sec:intro},
including Yakusu~\cite{fazzini2018automatically},  ReCDroid~\cite{zhao2019recdroid},  GIFdroid~\cite{feng2022gifdroid}, DroidScope~\cite{huang2023context},  ReproBot~\cite{zhang2023automatically}, AdbGPT~\cite{feng2024prompting}, CrashTranslator~\cite{huang2024crashtranslator}, Roam~\cite{zhang2024mobile} and ReBL~\cite{wang2024feedback}.  Section~\ref{tools} provides introductions to each of these approach.
These tools aim to use various types of information in bug reports—such as crash logs, full textual reports, and videos—to automate the bug reproduction process. 

However, none have considered the role of images, though GIFdroid automates bug replay from GIFs and videos in bug reports. While it involves processing GIFs and videos using image techniques, its goal is fundamentally different from our study. GIFdroid focuses on replaying sequential interactions from video-based bug reports, leveraging the temporal continuity of frames. In contrast, our work investigates how standalone images—without sequential context—can be utilized for automated bug reproduction. This distinction highlights the unique challenges and opportunities in leveraging images independently, motivating the need for our empirical study.

%

\noindent\textbf{Bug Report Study and Analysis.}
There have been several research efforts dedicated to studying and analyzing Android bug reports.
For instance, Wendland et al~\cite{wendland2021andror2} studied 90 manually reproduced bug reports to support automated research in bug analysis and reproduction. Building on this basis, Johnson et al~\cite{johnson2022empirical} extended the dataset and conducted an empirical study on 180 Android bug reports to examine reproduction challenges and the quality of reported details.
Chaparro et al.\cite{chaparroDetectingMissingInformation2017} conducted a study on user-reported behaviors, reproduction steps, and expected outcomes, identifying typical discourse patterns. They later developed Euler\cite{chaparroAssessingQualitySteps2019}, an automated method to evaluate the quality of S2R in Android bug reports based on simple grammar patterns. Liu et al.~\cite{liuAutomatedClassificationActions2020} introduced Maca, a machine-learning classifier that organizes S2R action words into standard categories.

Several studies have explored using LLMs to analyze and interpret bug reports. Lee et al.\cite{lee2022light} applied LLMs for bug triage, while Messaoud et al.\cite{messaoud2022duplicate} used a BERT model to detect duplicate reports. Kang et al.~\cite{kang2023large} proposed generating JUnit test methods for Java programs from bug reports

Some research focuses on improving the bug reporting process. Moran et al.\cite{moran2015auto} developed Fusion, which uses dynamic analysis to capture app UI events for more informative reports. Fazzini et al.\cite{fazzini2022enhancing} support reporters in writing precise reproduction steps by leveraging static and dynamic analysis to predict next actions. Yang et al.~\cite{song2022toward} offer a guided system with instant feedback and graphical hints to enhance report quality. 

However, these studies focus solely on improving the accuracy of identifying textual S2Rs, with no consideration given to images. Our study examines not only the functional role of images in automated bug report reproduction but also their characteristics within bug reports, offering insights into  potential opportunities to leverage them in automated bug reproduction.
\section{Conclusions}
\label{sec:conclusion}

In this paper, we conduct an empirical study on images in bug reports to reveal their characteristics and functions, analyzing their potential for use in automated bug report reproduction. We identify patterns in image quantity and types, examining their functional roles, usage patterns, and the essential information conveyed by each role. These insights lay the groundwork for future research aimed at differentiating images based on their roles and exploring potential techniques to leverage each role effectively in automated bug reproduction. This research also opens opportunities for enhancing existing tools or developing new, versatile tools that can integrate a broader range of information from bug reports to improve reproduction accuracy.

\section*{Acknowledgments}
This work was supported in part by the U.S. National
Science Foundation (NSF) under grants 
CCF-2402103, CCF-2403617, CCF-2403747, and CCF-2211454.

\bibliographystyle{IEEEtran}
\bibliography{paper}

\end{document}